\documentclass[aps,prl,twocolumn,a4paper,10pt,notitlepage,footinbib,showpacs]{revtex4}
\usepackage[english]{babel}
\usepackage[T1]{fontenc}
\usepackage{endnotes}
\usepackage{amssymb,amsmath,amsfonts}
\usepackage{textcomp}
\usepackage{graphics,color}

\begin{document}

\title{Activity-induced phase separation and self-assembly in mixtures\\of active and passive particles}

\author{Joakim Stenhammar}
\email{j.stenhammar@ed.ac.uk}
\affiliation{SUPA, School of Physics and Astronomy, University of Edinburgh, Edinburgh EH9 3FD, United Kingdom}

\author{Raphael Wittkowski}
\affiliation{SUPA, School of Physics and Astronomy, University of Edinburgh, Edinburgh EH9 3FD, United Kingdom}

\author{Davide Marenduzzo}
\affiliation{SUPA, School of Physics and Astronomy, University of Edinburgh, Edinburgh EH9 3FD, United Kingdom}

\author{Michael E. Cates}
\affiliation{SUPA, School of Physics and Astronomy, University of Edinburgh, Edinburgh EH9 3FD, United Kingdom}

\date{\today}
\pacs{82.70.Dd, 64.75.Xc, 64.75.Yz}
% 82.70.Dd: Colloids
% 64.75.Xc: Phase separation and segregation in colloidal systems
% 64.75.Yz: Self-assembly
% 83.10.Mj: Molecular dynamics, Brownian dynamics

\begin{abstract}
We investigate the phase behavior and kinetics of a monodisperse mixture of active (\textit{i.e.}, self-propelled) and passive isometric Brownian particles through Brownian dynamics simulations and theory. As in a purely active system, motility of the active component triggers phase separation into a dense and a dilute phase; in the dense phase we further find active-passive segregation, with ``rafts'' of passive particles in a ``sea'' of active particles. We find that phase separation from an initially disordered mixture can occur with as little as 15 percent of the particles being active. Finally, we show that a system prepared in a suitable fully segregated initial state reproducibly self-assembles an active ``corona'' which triggers crystallization of the passive core by initiating a compression wave. Our findings are relevant to the experimental pursuit of directed self-assembly using active particles. 
\end{abstract}

\maketitle

Understanding the collective behavior of systems composed of self-propelled (``active'') constituents is of great importance both from a fundamental physics perspective and for understanding many biological systems, such as bacterial suspensions, fish schools and bird flocks \cite{Marchetti-2013}. One example of such collective behavior is seen in so-called active Brownian particles -- self-propelled particles whose propulsion direction relaxes through rotational diffusion -- which have been shown to phase separate into a dense and a dilute phase even in the absence of attractive or aligning interactions \citep{Fily-2012,Redner-2013,Buttinoni-2013,Stenhammar-2013,Bialke-2013,Stenhammar-2014,Henkes-2014,WysockiWG2014,SpeckBML2014,WittkowskiTSAMC2014,Anisometric}. The driving force behind this phase separation is a positive accumulation feedback triggered by a slow-down of particles due to collisions, which can be represented by a propulsion speed $v(\rho)$ that decreases with the local particle density $\rho$ \cite{Tailleur-2008,Cates-2013,Stenhammar-2013}. The phase behavior of such systems is controlled by the total average particle packing fraction $\phi_0$ and the P\'eclet number $\mathrm{Pe} = 3v_0 \tau_{\mathrm{r}} / \sigma$, which essentially controls the ratio between motility and diffusion, where $v_0$ is the propulsion speed of an isolated particle, $\tau_{\mathrm{r}}$ its rotational relaxation time, and $\sigma$ its diameter.   

\begin{figure}[h!] 
\begin{center}
\resizebox{!}{80mm}{\includegraphics{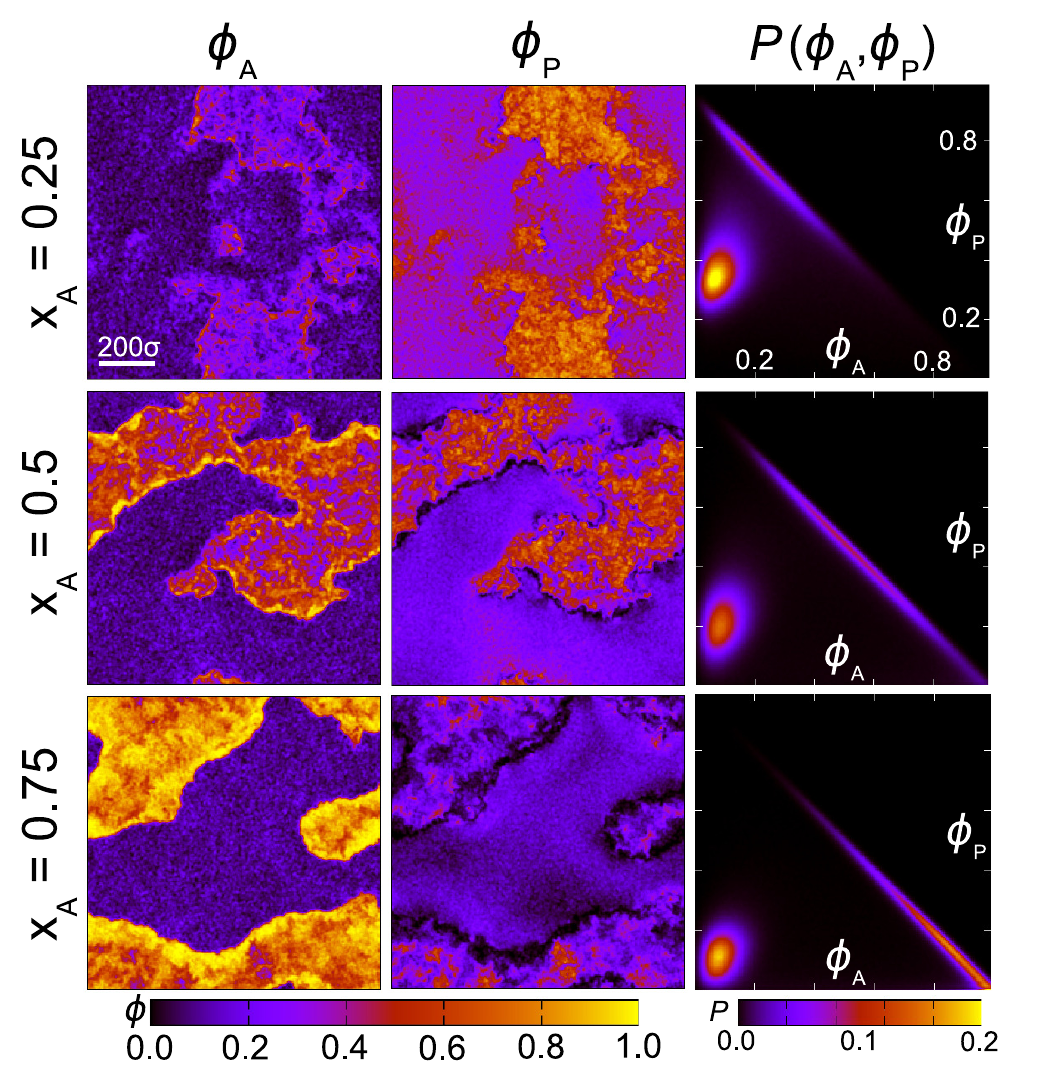}}
\caption{Left and center columns: Density plots of the local area fractions of active ($\phi_{\mathrm{A}}$) and passive ($\phi_{\mathrm{P}}$) particles for Pe = 300, $\phi_0 = 0.6$, and three different values of the fraction $x_{\mathrm{A}}$ of active particles, obtained at late times ($t = 5000 \tau_{\mathrm{LJ}}$) after a quench from a random initial configuration. 
The box size is $1000 \sigma \times 1000 \sigma$, corresponding to $N \approx 760 000$ particles. 
Right column: Two-dimensional probability distribution $P(\phi_{\mathrm{A}},\phi_{\mathrm{P}})$ obtained by averaging the corresponding simulations over the time window $4500 \tau_{\mathrm{LJ}} \leq t \leq 5000 \tau_{\mathrm{LJ}}$.} 
\label{snapshots}
\end{center}
\end{figure}

Experimentally, it is very challenging to reach the high packing fractions ($\phi_{0} \approx 0.5$) required for phase separation using only active particles \cite{Brown-2014}. Therefore, the phase-separation behavior of suspensions where the microswimmers are mixed with regular passive colloidal particles is of great practical interest. To date, most studies on active-passive mixtures have focused on the motion of individual passive tracer particles in swimmer suspensions \citep{Leptos-2009,Dunkel-2010,Mino-2013,Jepson-2013}. Only a small number of investigations have addressed the behavior of dense mixtures of active and passive agents; these have found interesting novel phenomena including active-passive segregation between rod-like particles \cite{McCandlish-2012}, crystallization of hard-sphere glasses \cite{Ni-2014}, emergence of flocking and turbulence \cite{Hinz-2014}, and a facilitation of attraction-induced phase separation \cite{Das-2014}. 

In this Letter, we provide a comprehensive numerical study of the complex phase behavior of mixtures of monodisperse isometric repulsive active Brownian particles and their passive counterparts in two dimensions. First, we show that a mixture prepared in a uniform phase is unstable to activity-induced phase separation, even when the fraction of active particles $x_{\mathrm{A}}$ is modest. We also find that the clusters formed are not homogeneous, but are predominantly active at the periphery and passive in their interior (see Fig.\ \ref{snapshots}). Compared to the purely active case, the dynamics of active-passive mixtures shows enhanced fluctuations, with frequent fission and fusion of clusters. Most strikingly, we show that, by carefully choosing our initial condition, the  mixture self-assembles into a disk-shaped core of passive particles surrounded by an active shell; the active particles further stimulate a compression wave which induces crystallization of the passive component. 

In our two-dimensional Brownian dynamics simulations, all particles (active as well as passive) are treated as monodisperse repulsive disks interacting through a truncated and shifted Lennard-Jones potential $U(r) = 4\varepsilon [ \left( \sigma / r \right)^{12} - \left( \sigma / r \right)^{6} ] + \varepsilon$ with an upper cut-off at $r = 2^{1/6}\sigma$, beyond which $U = 0$. Here, $\varepsilon$ determines the interaction strength as well as the Lennard-Jones time scale $\tau_{\mathrm{LJ}}=\sigma^{2}/(\varepsilon\beta D_{\mathrm{t}})$; $r$ is the center-to-center distance between two particles, and $\beta = 1/(k_{\mathrm{B}}T)$ is the inverse thermal energy. We studied this model by solving the overdamped Langevin equations (thus neglecting hydrodynamic interactions between particles)
\begin{align}
\partial_t \mathbf{r}_i &= \beta D_{\mathrm{t}} \left( \mathbf{F}_i + F_{\mathrm{A}} \mathbf{p}_i \right) + \sqrt{2D_{\mathrm{t}}} \:\! \boldsymbol{\Lambda}_{\mathrm{t}} \;, \label{Langevin_t} \\
\partial_t \theta_i &= \sqrt{2D_{\mathrm{r}}} \:\! \Lambda_{\mathrm{r}} \;, 
\label{Langevin_r_2D}%
\end{align}
where $\mathbf{r}_{i}(t)$ is the position and $\theta_{i}(t)$ is the orientation of the $i$th particle at time $t$ \cite{FurtherDetails}, in a box with periodic boundary conditions.  
$\mathbf{F}_i$ is the total conservative force on particle $i$, which results from $U(r)$, $F_{\mathrm{A}}$ is the constant magnitude of the self-propulsion force $F_{\mathrm{A}}\mathbf{p}_{i}$ on particle $i$ ($F_{\mathrm{A}} = 24 \varepsilon/\sigma$ for active and $F_{\mathrm{A}} = 0$ for passive particles, see \cite{Stenhammar-2014} for details), and $\mathbf{p}_i = (\cos \theta_i, \sin \theta_i)$ its direction. 
Furthermore, $D_{\mathrm{t}}$ and $D_{\mathrm{r}} = 1/\tau_{\mathrm{r}} = 3D_{\mathrm{t}}/\sigma^{2}$ denote the translational and rotational diffusion coefficients of the particles, respectively; $\boldsymbol{\Lambda}_{\mathrm{t}}(t)$ and $\Lambda_{\mathrm{r}}(t)$ are unit-variance Gaussian white noise terms 
\cite{Noise}.

Figure \ref{snapshots} (see also movies in \cite{SI}) shows the 
late-time local area fractions $\phi_{\mathrm{A}}$ and $\phi_{\mathrm{P}}$ of active ($\phi_{\mathrm{A}}$) and passive ($\phi_{\mathrm{P}}$) particles in initially homogeneous mixtures with $N\approx 760000$ particles as well as the corresponding probability distributions $P(\phi_{\mathrm{A}},\phi_{\mathrm{P}})$ 
for active-particle fractions $x_{\mathrm{A}}\in\{0.25,0.5,0.75\}$. 
For all $x_{\mathrm{A}}$ the mixture phase separates into a dense ``liquid'' phase and a dilute ``gas'' phase. The probability plots also suggest that (\textit{i}) the area fraction $\phi_{\mathrm{A,g}}$ of active particles in the gas phase remains essentially constant as $x_{\mathrm{A}}$ is varied (unlike the corresponding passive area fraction $\phi_{\mathrm{P,g}}$ which varies markedly), and (\textit{ii}) the \textit{total} area fraction $\phi_{\mathrm{A,l}} + \phi_{\mathrm{P,l}}$ in the liquid phase remains constant and close to unity as indicated by the diagonal straight lines in $P(\phi_{\mathrm{A}},\phi_{\mathrm{P}})$. 

From the snapshots in Fig.\ \ref{snapshots} it is also apparent that the distribution of active and passive particles within the dense phase is inhomogeneous, with the active particles being significantly enriched at the boundaries of the dense domains. The interfacial layer remains essentially purely active for all $x_{\mathrm{A}}$. Since the flux balance between this layer and the gas determines $\phi_{\mathrm{A,g}}$, this corroborates our finding that $\phi_{\mathrm{A,g}}$ remains constant as $x_{\mathrm{A}}$ is varied.
However, even in the interior of the dense phase, there is clearly an inhomogeneous distribution of the two species, with the passive particles agglomerating into domains.  
As can be seen from Fig.\ S1 and movies in \cite{SI}, the domain size of these ``passive rafts'' inside the ``active sea'' initially increases with time, and eventually reaches a (noisy) steady-state plateau value of approximately $20 \sigma$. Such spontaneous segregation between otherwise identical active and passive particles has only been observed before in the context of rod-like particles \cite{McCandlish-2012}; our results suggest that it is the result of activity alone and occurs also for mixtures of isometric particles (see further \cite{SI}). This phenomenon may be seen as a manifestation, at a larger density and with equal-sized particles, of the effective attraction observed between passive colloids \citep{Angelani-2011} and between hard walls \cite{Bolhuis-2014,Reichhardt-2014} in a bacterial bath. 

We also note that the overall dynamics of active-passive mixtures is very different from that of the corresponding purely active systems \cite{Stenhammar-2013} (see movies in \cite{SI}). Most importantly, the dynamics of phase-separating mixtures is much more violent, with clusters constantly moving, fissioning, and merging. 

\begin{figure}[ht] 
\begin{center}
\resizebox{!}{40mm}{\includegraphics{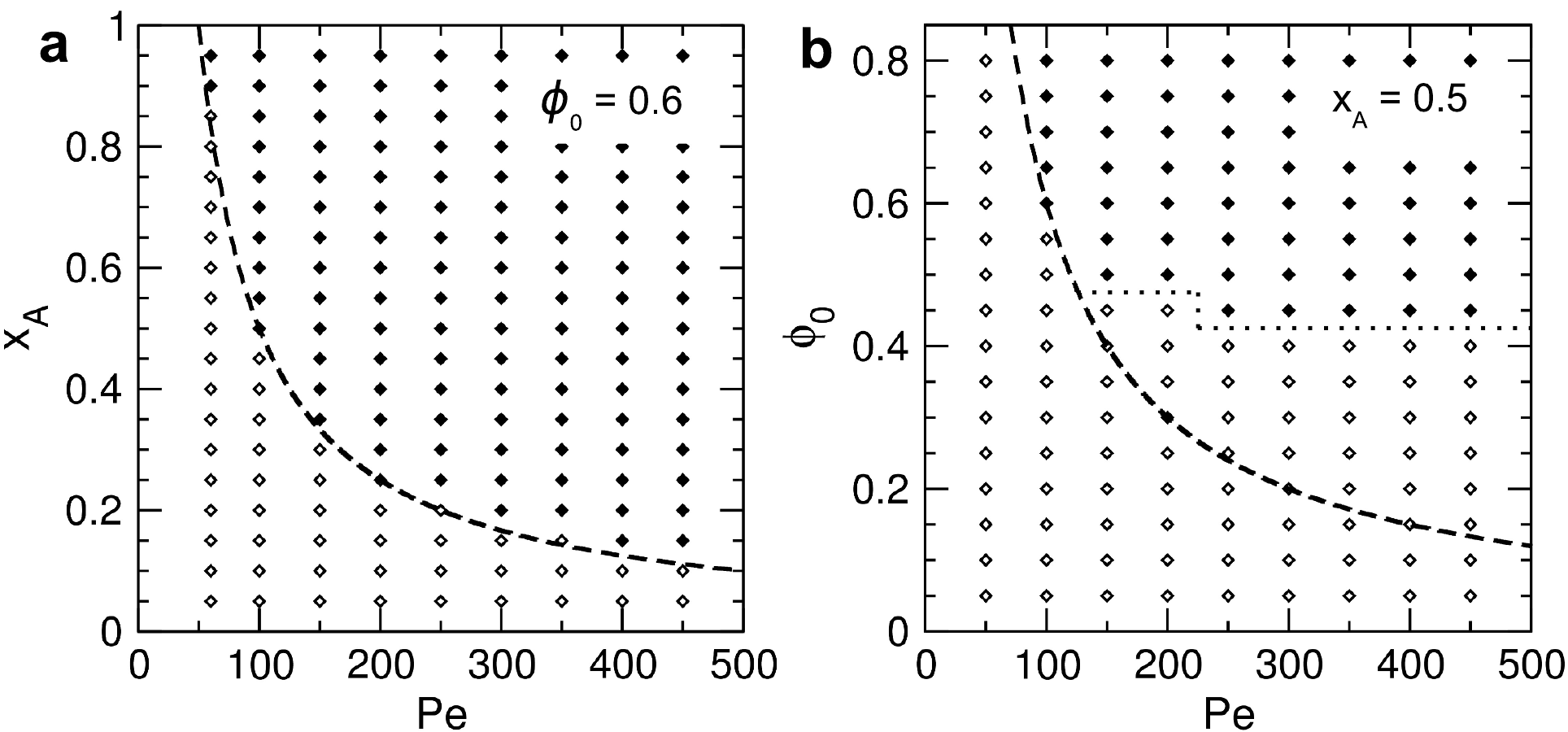}}
\caption{(a) Phase diagram in the $\mathrm{Pe}$-$x_{\mathrm{A}}$ plane for $\phi_0 = 0.6$. 
(b) Phase diagram in the $\mathrm{Pe}$-$\phi_0$ plane for $x_{\mathrm{A}} = 0.5$. 
Filled symbols denote phase-separated systems, as determined by visual inspection, and open symbols denote homogeneous systems. All results were obtained by a quench from random initial configurations, using systems of size $150\sigma \times 150\sigma$, corresponding to $N \approx 15000$ particles for $\phi_0 = 0.6$. The dashed lines indicate phase boundaries predicted by Eq.\ \eqref{binodal} with $\kappa = 4.05$.}
\label{phase_diagrams}%
\end{center}
\end{figure}

Figure \ref{phase_diagrams} shows phase diagrams in the Pe-$x_{\mathrm{A}}$ and Pe-$\phi_0$ planes, where $\phi_0$ is the total particle area fraction and $0 \leq x_{\mathrm{A}} \leq 1$ is the fraction of the particles that are active. Interestingly, spontaneous phase separation occurs with $x_{\mathrm{A}}$ as small as 0.15, or equivalently with $\phi_{\mathrm{A}}=0.09$ (see Fig.\ \ref{phase_diagrams}a). Such area fractions should be easily achievable in experiments, although the corresponding P\'eclet number required for phase separation is relatively high ($\approx$500 according to our simulations, as opposed to $\approx$60 for a purely active system \cite{Stenhammar-2013}).

Figure \ref{phase_diagrams}a also shows that the phase boundary between the uniform and the phase-separated region closely follows a fit to the function $x_{\mathrm{A}} \sim 1/\mathrm{Pe}$ (dashed line). To understand this behavior, we start from the observations of Fig.\ \ref{snapshots} that $\phi_{\mathrm{A,g}}$ is approximately independent of $x_{\mathrm{A}}$ and that the total density of the dense phase is close to the close-packing density $\phi_{\mathrm{cp}} = \pi / (2\sqrt{3}) \approx 0.907$ (this is also true for the corresponding purely active system \cite{Redner-2013}). Following the kinetic model developed by Redner \textit{et al.} in Ref.\  \cite{Redner-2013}, we equate the incoming and outgoing fluxes $k_{\mathrm{in}}$ and $k_{\mathrm{out}}$ of active particles at the interface of a dense cluster. Using $k_{\mathrm{in}} = 4 \phi_{\mathrm{A,g}} v_0 /(\pi^2 \sigma^2)$ and $k_{\mathrm{out}} = \kappa D_{\mathrm{r}} / \sigma$, where $\kappa$ is a dimensionless parameter introduced to take into account the fact that particles leave the dense phase in bursts \cite{Redner-2013}, yields $\phi_{\mathrm{A,g}} = 3\pi^2\kappa / (4\mathrm{Pe})$. In the purely active case, the binodal line is given by the relationship $\phi_0 = \phi_{\mathrm{A,g}}$ \cite{Redner-2013}. In the case of active-passive mixtures, the condition is instead $\phi_0 x_{\mathrm{A}} = \phi_{\mathrm{A,g}}$, leading to the following expression for the binodal:
\begin{equation}
\phi_0 x_{\mathrm{A}} = \frac{3\pi^2\kappa}{4\mathrm{Pe}} \;.
\label{binodal}%
\end{equation}
The dashed lines in Fig.\ \ref{phase_diagrams} show Eq.\ \eqref{binodal} with $\kappa \approx 4.05$, close to the value found for $x_{\mathrm{A}}=1$ \cite{Redner-2013}. For the case of constant $\phi_0$ and varying $x_{\mathrm{A}}$ (see Fig.\ \ref{phase_diagrams}a), the agreement is excellent, showing that the basic kinetic assumptions made above (and originally proposed to describe a purely active system) provide a workable model for active-passive mixtures with $x_{\mathrm{A}}$ as low as 0.15, and for a significantly larger range of P\'eclet numbers than studied previously \cite{Redner-2013}. For the case of constant $x_{\mathrm{A}} = 0.5$ and varying $\phi_0$, the agreement between simulation results and Eq.\ \eqref{binodal} is good down to $\phi_0 \approx 0.45$, below which the simulated systems stop phase separating altogether (see Fig.\ \ref{phase_diagrams}b). This discrepancy arises because the approximations inherent in Eq.\ \eqref{binodal} are known to be inappropriate as $\mathrm{Pe} \rightarrow \infty$, where phase separation does not occur if the decrease of $v(\rho)$ with $\rho$ is not steep enough \cite{Henkes-2014,Tailleur-2008,Cates-2013,Stenhammar-2013}. This also explains why in practice there exists a lower value of $x_{\mathrm{A}}$ required for spontaneous phase separation, while in principle Eq.\ \eqref{binodal} predicts phase separation to occur even for $x_{\mathrm{A}} \rightarrow 0$ as $\mathrm{Pe} \rightarrow \infty$.

To describe this lower ``spinodal line'' it is therefore necessary to extend the continuum theories developed in Refs.\ \cite{Cates-2013,Bialke-2013} towards active-passive mixtures. While we postpone the detailed treatment of this problem to a future study \cite{WittkowskiEtAl}, we find that the simple assumption of a linear dependence of the propulsion speed $v$ on each of the area fractions $\phi_{\mathrm{A}}$ and $\phi_{\mathrm{P}}$, \textit{i.e.}, $v = v_0 (1-a\phi_{\mathrm{A}}-b\phi_{\mathrm{P}})$ with constant parameters $a$ and $b$, together with the assumption that $\phi_{\mathrm{P}}$ remains uniform during the initial spinodal instability leads to the spinodal condition $\phi_0 > 1/[2a x_{\mathrm{A}} + b(1-x_{\mathrm{A}})]$ \cite{SpinodalCondition}. Simulations in the one-phase region of the phase diagram confirm the linear dependence of $v$ on both area fractions and yield the approximate values $a = 1.08$ and $b = 1.21$ \cite{SI}. For $x_{\mathrm{A}} = 0.5$, this gives a spinodal density $\phi_0 \approx 0.59$, in reasonable agreement with the value $\phi_0 = 0.45-0.50$ observed in Fig.\ \ref{phase_diagrams}b; the slight mismatch can be attributed to the fact that spontaneous phase separation will occur slightly outside the spinodal due to the high noise level. 

\begin{figure}[ht] 
\begin{center}
\resizebox{!}{42mm}{\includegraphics{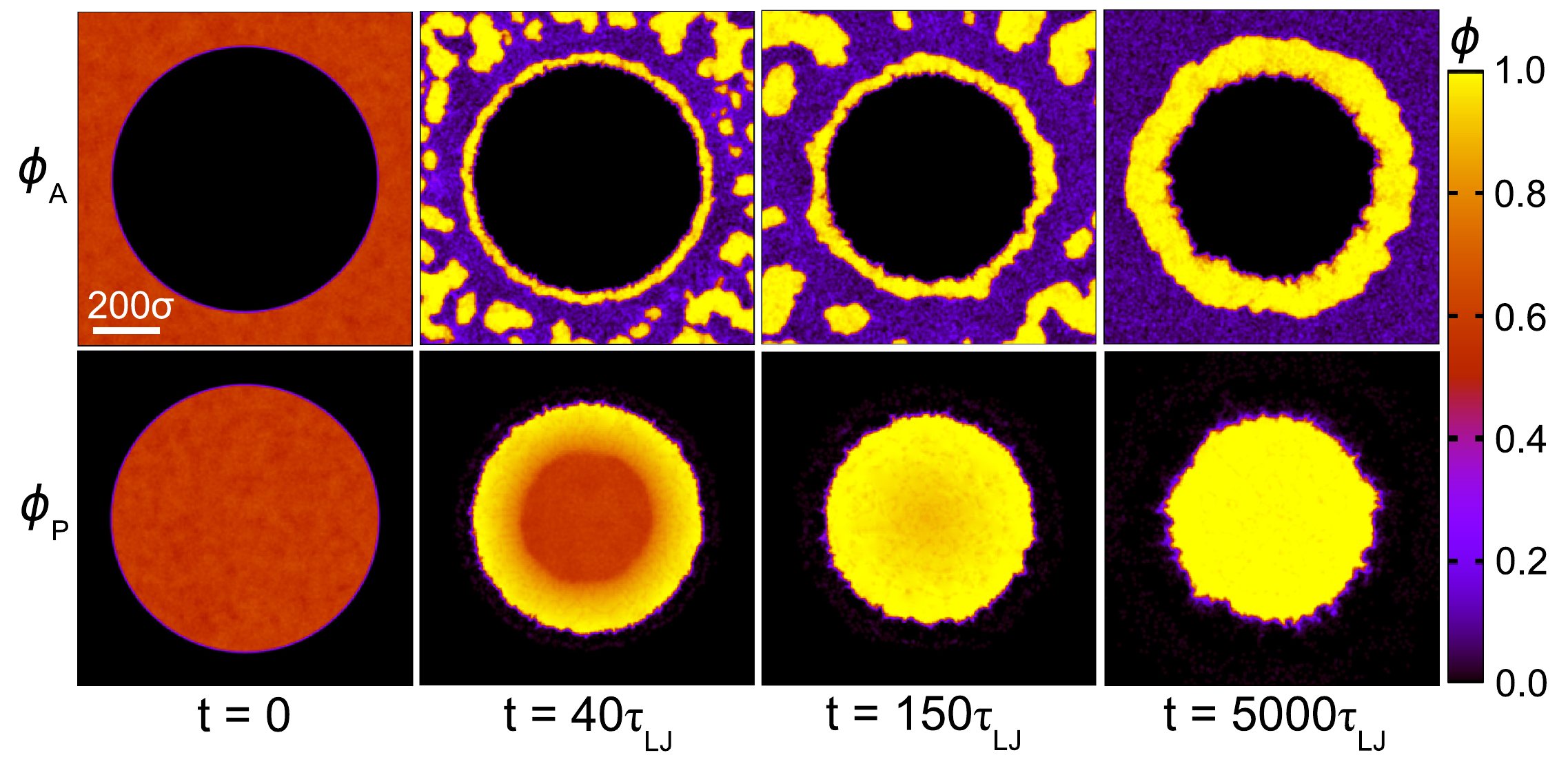}}
\caption{Density plots showing the local area fractions of active ($\phi_{\mathrm{A}}$) and passive ($\phi_{\mathrm{P}}$) particles at different times after a quench, starting with a fully segregated initial configuration. The system parameters are as in Fig.\ \ref{snapshots} with $x_{\mathrm{A}} = 0.5$.}
\label{snapshots_sphere}%
\end{center}
\end{figure}

\begin{figure}[ht] 
\begin{center}
\resizebox{!}{100mm}{\includegraphics{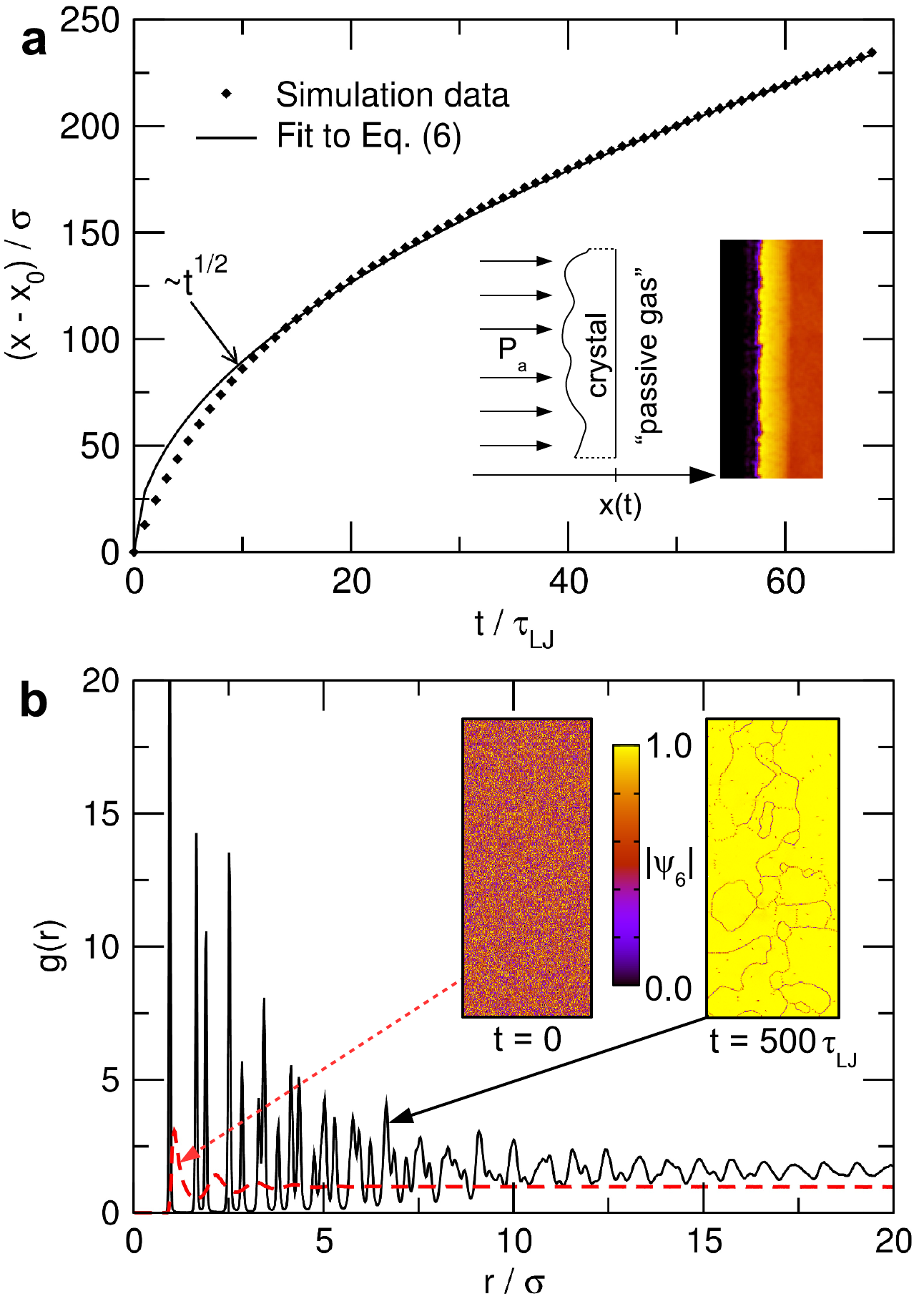}}
\caption{(a) Time-dependent position $x(t)-x_{0}$ of the shockwave front traveling through a slab of passive particles after a quench, for $x_{\mathrm{A}} = 0.5$. $x(t)$ was measured by averaging the density in the $y$-direction to create a one-dimensional density profile with a well-defined wavefront. The inset shows the thin crystalline layer which forms at the wavefront. (b) Pair-distribution function $g(r)$ of the passive particles before (red dashed line) and long after (black solid line) the quench. The inset shows the magnitude of the local hexagonal order parameter $\psi_{6}(\mathbf{r}_i) = N_{i}^{-1} \sum^{N_{i}}_{j=1} \exp(6\mathrm{i}\theta_{ij})$ in the interior of the passive phase, where $\mathrm{i}$ is the imaginary unit, $\theta_{ij}$ is the angle between an arbitrary reference axis and the displacement vector between particles $i$ and $j$, and the sum runs over all $N_{i}$ particles within a cut-off radius of $1.3 \sigma$ from particle $i$.} 
\label{shockwave}%
\end{center}
\end{figure}

Figures \ref{snapshots} and S1 \citep{SI} clearly indicate that there is a tendency towards segregation between active and passive isometric particles. Starting from this premise, and with a view towards speeding up pattern formation, we set out to study the behavior of a system initialized from a state where active and passive particles are fully segregated with the passive particles forming a disk-shaped cluster surrounded by the active particles (see Fig.\ \ref{snapshots_sphere} and movies in \cite{SI}).  
With this initial state, we find a very different dynamics with respect to the one previously analyzed (compare Figs.\ \ref{snapshots} and \ref{snapshots_sphere}). Now, the passive particles are quickly pushed together by a pressure wave which starts at the active-passive boundary and travels inward into the passive phase (see also Fig.\ \ref{shockwave}). After the wave has reached the center of the cluster, there is a second relaxation period, resulting in a very dense phase consisting almost exclusively of passive particles. Thereafter, the active particles gradually condense onto the interface, creating a ``corona'' of active dense phase, whose thickness self-adjusts so as to yield the same value of $\phi_{\mathrm{A,g}}$ as seen for the random initial configuration (see Fig.\ \ref{snapshots}). This final, compressed configuration remains stable for the whole duration of the simulation (corresponding to $\approx$400 characteristic diffusion times $\tau \equiv \sigma^2/D_{\mathrm{t}}$, or $\approx$900 seconds in a system of micron-sized spherical colloids), in contrast to the much more violent dynamics observed throughout the simulation for random initial configurations. Simulations of smaller systems \cite{SI}, however, indicate that the passive domain will eventually dissolve and the system will reach the same mesoscopically segregated steady state as seen when starting from random initial conditions. 

To further analyze the compression of the passive phase driven by the active component, in Fig.\ \ref{shockwave}a we present an analysis of the travelling pressure wavefront, obtained from a simulation started in a fully segregated slab geometry in order to facilitate the analysis. The wavefront position $x(t)$ is well described by $(x - x_0) \sim t^{1/2}$, where $x_0=x(0)$ is the initial position of the active-passive interface. To understand this scaling, we assume that the effect of the active component on the passive one can be lumped into a constant two-dimensional ``active pressure'' $P_{\mathrm{a}}$ \cite{ActivePressure,Fily-2014,Yang-2014,Brady-2014}. As our dynamics is overdamped, this active pressure will lead to the propagation of a compression wave, whose velocity $\dot{x}(t)$ at the wavefront can be estimated as
\begin{equation}
\dot{x}(t) = \frac{\rho_{\mathrm{l}}}{\rho_{\mathrm{l}}-\rho_{\mathrm{g}}} \frac{F_{\mathrm{a}}}{M(t) \gamma} \;,
\label{x_dot}%
\end{equation}
where $F_{\mathrm{a}} = L P_{\mathrm{a}}$ is the active force acting on a slab segment of length $L$, $M(t)$ is the total mass of the particles carried along by this segment of the wavefront, $\gamma$ is the damping rate (in units of inverse time) of the implicit solvent, and $\rho_{\mathrm{l}}$ and $\rho_{\mathrm{g}}$ denote the number densities of the dense and dilute phases, respectively. The prefactor involving the densities takes into account the relative velocities of the front and the back of the wave, and can be derived through a straightforward mass conservation argument \citep{SI}. 

We now note that the total mass $M$ of the slab segment is proportional to the total area swept by the wavefront, $M(t) = m \rho_{\mathrm{g}} L (x(t)-x_0)$, where $m$ is the mass of a single particle. Inserting this expression into Eq.\ \eqref{x_dot} yields the differential equation 
\begin{equation}
\dot{x} = \frac{P_{\mathrm{a}}\rho_{\mathrm{l}}}{m\gamma  \rho_{\mathrm{g}} (\rho_{\mathrm{l}}-\rho_{\mathrm{g}}) (x-x_0)} \;, 
\end{equation}
with the solution  
\begin{equation}\label{x_t}
x(t) = x_{0} + \left( \frac{2P_{\mathrm{a}} \rho_{\mathrm{l}} }{m \gamma \rho_{\mathrm{g}} (\rho_{\mathrm{l}}-\rho_{\mathrm{g}})} \right)^{1/2} t^{1/2} \;.
\end{equation}
The fitting parameter $P_{\mathrm{a}}$ for the solid line in Fig.\ \ref{shockwave}a corresponds to $P_{\mathrm{a}} \approx 120 \varepsilon / \sigma^2$. With the single-particle propulsion force $F_{\mathrm{a}} = 24 \varepsilon / \sigma$ this means that the pressure during the compression is comparable to about 5 layers of active particles compressing the slab, in broad agreement with what we observe in the simulations (see movies in \cite{SI}). Finally, the pair distribution functions and hexagonal order parameters shown in Fig.\ \ref{shockwave}b clearly show that the final state of the two-dimensional passive suspension is a crystalline one. These results suggest that judicious mixing with active particles can be used as a tool to control self-assembly of passive colloids, at least in two dimensions.

In this Letter, we have presented a systematic study of mixtures of active and passive Brownian particles with varying composition and density. Apart from being of fundamental interest from a non-equilibrium physics perspective, understanding such mixtures is a prerequisite for experiments intended to create novel materials through active phase separation and self-assembly. We have shown that activity-induced phase separation is indeed possible for a wide range of system parameters, and for active-to-passive ratios as small as 1:6, as long as the P\'eclet number is large enough. This is encouraging from an experimental viewpoint given the difficulties associated with the manufacturing and fuel supply of large quantities of active colloids ~\citep{Brown-2014}. We have further shown that the choice of appropriate initial conditions, where the two species are segregated, leads to a remarkable directed assembly process, whereby the active particles drive a compression wave through the passive phase which leads to the creation of a passive colloidal crystal. The active particles then coalesce and form a highly fluctuating wetting layer around the crystal. These results call for a future experimental exploration of active-passive mixtures and open up new potential routes to directed assembly using active particles.

\begin{acknowledgments}
We thank Rosalind Allen for input and helpful discussions, and EPSRC (Grant No.\ EP/J007404) for funding. 
J.S. gratefully acknowledges financial support from the Swedish Research Council (Grant No.\ 350-2012-274), 
R.W. gratefully acknowledges financial support through a Postdoctoral Research Fellowship (Grant No.\ WI 4170/1-2) from the German Research Foundation (DFG), and 
M.E.C. through a Royal Society Research Professorship. 
\end{acknowledgments}

\appendix

\subsection*{A. Active-passive segregation in the dense phase}
\begin{figure}[h] 
\begin{center}
\resizebox{!}{100mm}{\includegraphics{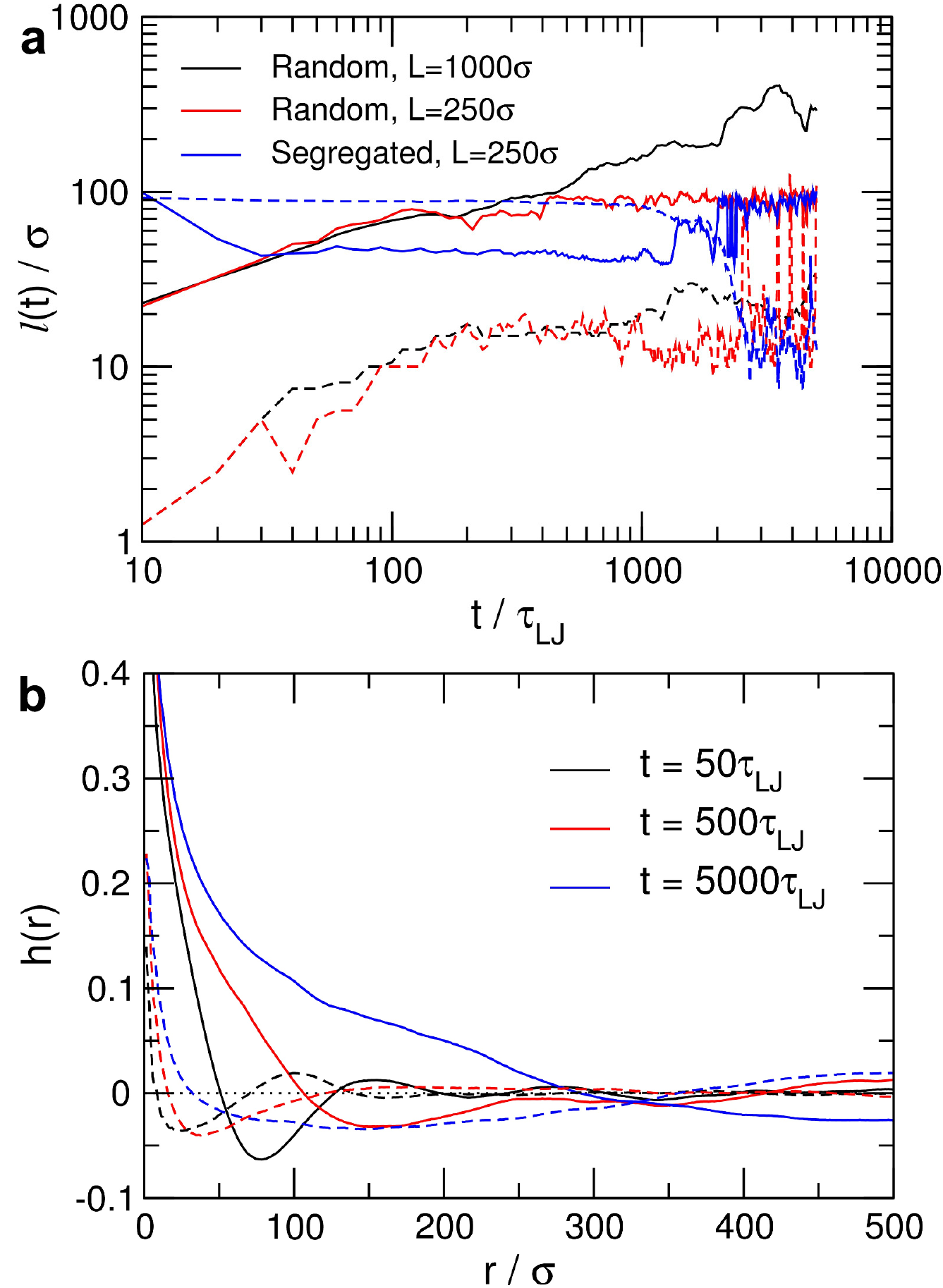}}
\caption{(a) Characteristic segregation length scales $l_{\mathrm{A}}(t)$ (solid lines) and $l_{\mathrm{P}}(t)$ (dashed lines) of dense active and passive domains, respectively, obtained from the first zero crossings of $h_{\mathrm{A}}(r)$ and $h_{\mathrm{P}}(r)$, for different lengths $L$ of the quadratic simulation box and for varying initial conditions. All results were obtained for $\mathrm{Pe} = 300$, $\phi_0 = 0.6$, and $x_{\mathrm{A}} = 0.5$. (b) $h_{\mathrm{A}}(r)$ (solid lines) and $h_{\mathrm{P}}(r)$ (dashed lines) for different times after a quench with $L=1000\sigma$ and random initial conditions.}
\label{L_t}%
\end{center}
\end{figure}

In Fig.\ \ref{L_t}a, the time-evolution of the characteristic length scales $l_{\mathrm{A}}(t)$ and $l_{\mathrm{P}}(t)$ of active ($l_{\mathrm{A}}$) and passive ($l_{\mathrm{P}}$) domains in the dense phase is shown. These length scales are measured from the locations of the first zero crossings of the functions $h_{\mathrm{A}}(r) = g_{\mathrm{AA}}(r) / g_{\mathrm{AP}}(r)-1$ and $h_{\mathrm{P}}(r) = g_{\mathrm{PP}}(r) / g_{\mathrm{AP}}(r)-1$, respectively, where $g_{\mu\nu}(r)$ is the radial pair-distribution function between species $\mu$ and $\nu$ (see Fig.\ \ref{L_t}b). Starting from a homogeneous system with a random initial distribution, the characteristic length scale of active domains exhibits a clear growth throughout the simulation, until it reaches a value comparable to the length of the simulation box. In contrast, the length-scale of passive domains shows an initial increase, indicating a gradual segregation between active and passive particles, until it reaches a plateau value of $20-30\sigma$, seemingly independent of the system size. This steady-state value is identical within error to the one observed in the smaller system initiated from a fully segregated disk-shaped state (blue dashed line in Fig.\ \ref{L_t}a).

\begin{figure}[h] 
\begin{center}
\resizebox{!}{50mm}{\includegraphics{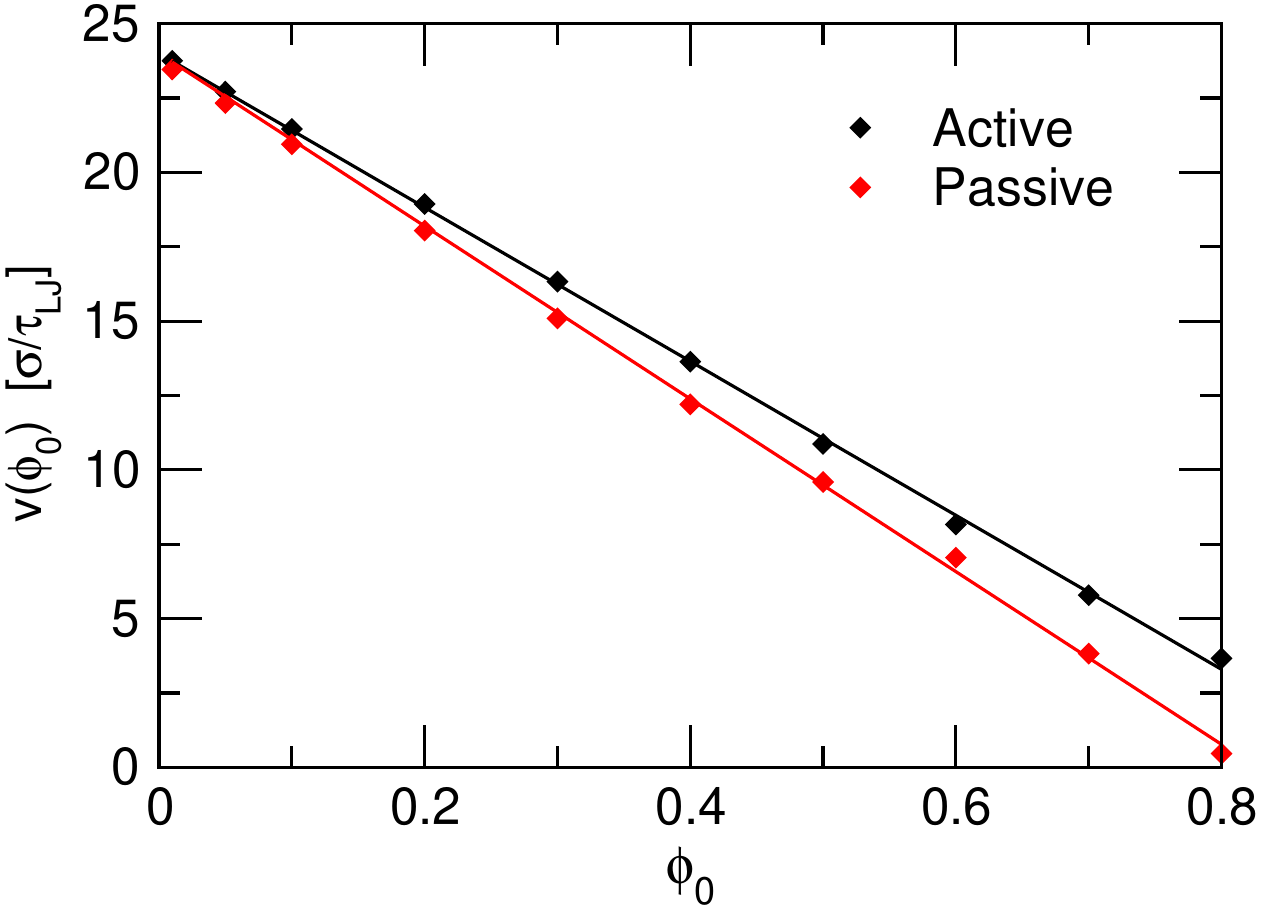}}
\caption{Effective propulsion speed $v$ of active particles in a suspension of active or passive particles as a function of the total particle area fraction $\phi_0$. In the passive case, $v$ was measured for a very small number ($N = 286$) of active particles, while $\phi_0$ was varied by adding passive particles. The solid lines are linear fits to the simulation data, given by $v(\phi_0) = v_0(1-1.08\phi_0)$ (active particles, black solid line) and $v(\phi_0) = v_0(1-1.21\phi_0)$ (passive particles, red solid line). All results were obtained for Pe = 50, \emph{i.e.}, in the one-phase region of the phase diagram. The dimensions of the systems were $150\sigma\times150\sigma$ and $v$ was measured as described in Refs.\ \citep{Stenhammar-2013,Stenhammar-2014}.}
\label{v_phi}%
\end{center}
\end{figure}

\subsection*{B. Spinodal condition}
For a one-component system of active particles it has been shown that a spinodal instability occurs if $\mathrm{d}\ln(v(\phi))/\mathrm{d}\ln(\phi) <-1$, where $v(\phi)$ is the density-dependent propulsion speed of the active particles \cite{Cates-2013}. For purely repulsive particles without hydrodynamic interactions, such as the ones studied here, $v(\phi) = v_0(1-\alpha\phi)$ is a linearly decreasing function of the packing fraction $\phi$ due to collisions, where $v_0$ is the propulsion speed of a free particle and $\alpha$ is a constant \citep{Stenhammar-2013,Bialke-2013,Henkes-2014}.

Assuming that the passive particles in a mixture of active and passive particles initially remain homogeneously distributed when the mixture becomes unstable, \emph{i.e.}, assuming that the instability of a mixture is triggered by a clustering of the active particles alone, a modified version of the spinodal condition for a one-component system of active particles can be used to describe the instability of an active-passive mixture. For this purpose we replace the density-dependent propulsion speed $v(\phi)$ of the one-component system with the density-dependent propulsion speed $v(\phi_{\mathrm{A}},\phi_{\mathrm{P}})=v_0 (1-a\phi_{\mathrm{A}}-b\phi_{\mathrm{P}})$ of a mixture (see Fig.\ \ref{v_phi}), and the total derivative with respect to $\ln(\phi)$ by a partial derivative with respect to $\ln(\phi_{\mathrm{A}})$:
\begin{equation}
\frac{\partial\ln(v(\phi_{\mathrm{A}},\phi_{\mathrm{P}}))}{\partial\ln(\phi_{\mathrm{A}})} < -1 \;.
\label{eq:ln_condition_mixture}%
\end{equation}%
With the total area fraction $\phi_0=\phi_{\mathrm{A}}+\phi_{\mathrm{P}}$ and the fraction of active particles $x_{\mathrm{A}}=\phi_{\mathrm{A}}/\phi_0$, Eq.\ \eqref{eq:ln_condition_mixture} leads to the following spinodal condition for active-passive mixtures:
\begin{equation}
\phi_0 > \frac{1}{2a x_{\mathrm{A}}+b(1-x_{\mathrm{A}})} \;.
\label{eq:spinodal_condition_mixture}%
\end{equation}
A more detailed and fully microscopic derivation will be given in a forthcoming publication \cite{WittkowskiEtAl}.

\subsection*{C. Velocity of the wavefront}
\begin{figure}[h] 
\begin{center}
\resizebox{!}{50mm}{\includegraphics{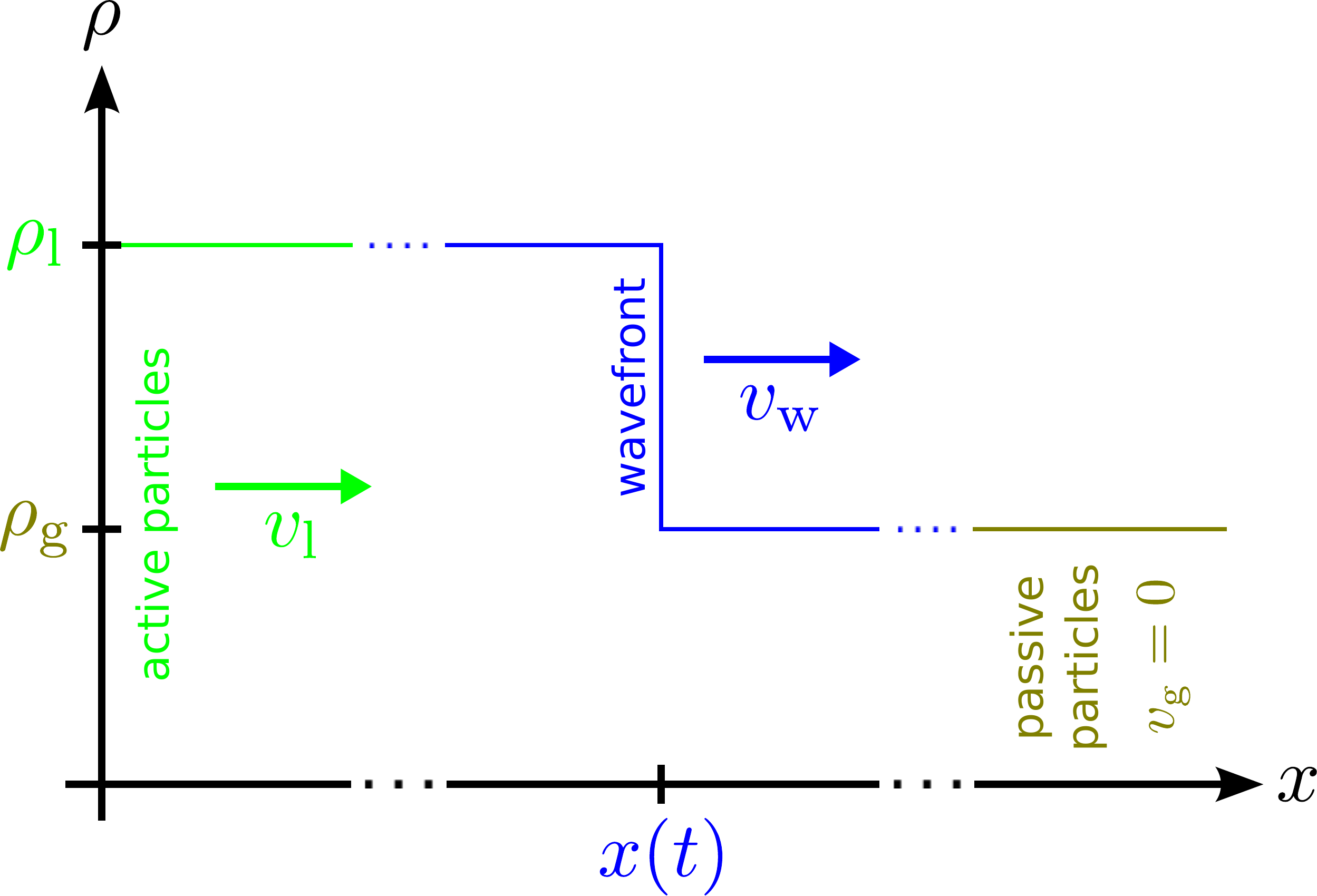}}
\caption{Sketch of the density profile $\rho(x)$ behind ($x<x(t)$) and in front of ($x>x(t)$) the compression wavefront ($x=x(t)$). Particle number conservation implies the relation $v_{\mathrm{l}}\rho_{\mathrm{l}}=v_{\mathrm{w}}(\rho_{\mathrm{l}}-\rho_{\mathrm{g}})$.}
\label{wavefront}%
\end{center}
\end{figure}

The density-dependent prefactor $\rho_{\mathrm{l}}/(\rho_{\mathrm{l}}-\rho_{\mathrm{g}})$ in Eq.\ (4) in the main text is a simple approximation for the ratio $v_{\mathrm{w}}/v_{\mathrm{l}}$ of the velocities $v_{\mathrm{w}}$ and $v_{\mathrm{l}}$ of the wavefront and the dense phase at the back of the compression wave, respectively (see Fig.\ \ref{wavefront}). Using the fact that the particles of the ``passive gas'' in front of the compression wave are stationary ($v_{\mathrm{g}}=0$), and equating the expressions for the particle fluxes at the front and the back of the wave results in the relation $v_{\mathrm{l}}\rho_{\mathrm{l}} = v_{\mathrm{w}}(\rho_{\mathrm{l}}-\rho_{\mathrm{g}})$. Together with $v_{\mathrm{l}} = F_{\mathrm{a}}/(M(t)\gamma)$ and $v_{\mathrm{w}} = \dot{x}(t)$ this leads to Eq.\ (4) in the main text. 

%$\rho_{\mathrm{l}}$ denotes the density and $v_{\mathrm{l}}$ the velocity of the dense ``liquid'' phase of the active particles, which propels the wavefront, and $\rho_{\mathrm{g}}$ is the density of the dilute ``gas'' phase of the passive particles before compression. 
%Since the passive particles in front of the wavefront have not yet been reached by the compression wave, their velocity $v_{\mathrm{g}}=0$ vanishes. 
%The simple approximation $\rho_{\mathrm{l}}/(\rho_{\mathrm{l}}-\rho_{\mathrm{g}})= v_{\mathrm{w}}/v_{\mathrm{l}}$ follows directly from particle number conservation (see Fig.\ \ref{wavefront}). 
%

\bibliography{bibliography}

\begin{thebibliography}{37}
\bibitem[{Noi()}]{Noise}
\bibinfo{howpublished}{We use the same noise terms for the active particles as
  for the passive ones, which is a common approximation is active-particle
  simulations \cite{Redner-2013,Stenhammar-2013}.}
\end{thebibliography}

\end{document}